\documentclass[pdflatex,sn-basic]{sn-jnl}
\usepackage{amsmath}
\usepackage{amsfonts}
\usepackage{amssymb}
\usepackage{lineno}
\usepackage{bigints}
\usepackage{graphicx}
\usepackage{color}
\usepackage[T1]{fontenc}
\usepackage{multirow}
\usepackage{booktabs}
\usepackage{bbm}
\usepackage{epstopdf}
\usepackage{placeins}

\DeclareMathOperator*{\argmax}{arg\,max}

\begin{document}

\title[Modeling and estimating skewed and heavy-tailed populations via unsupervised mixture models]{Modeling and estimating skewed and heavy-tailed populations via unsupervised mixture models}

\author*[1]{\fnm{Marco} \sur{Bee}}\email{marco.bee@unitn.it}

\author[2]{\fnm{Flavio} \sur{Santi}}\email{flavio.santi@unitn.it}

\affil*[1]{\orgdiv{Department of Economics and Management}, \orgname{University of Trento}, \orgaddress{\street{via Inama, 5}, \city{Trento}, \postcode{38122}, \country{Italy}}}

\affil[2]{\orgdiv{Department of Economics and Management}, \orgname{University of Trento}, \orgaddress{\street{via Inama, 5}, \city{Trento}, \postcode{38122}, \country{Italy}}}


\abstract{
We develop a mixture model for non-negative, heavy-tailed data, such as losses in actuarial and risk management applications. The mixture has a lognormal component, which is usually appropriate for the body of the distribution, and a Pareto-type tail, aimed at accommodating the largest observations, since the lognormal often decays too fast. Given that the tail is modeled by a zero-location Generalized Pareto distribution, the model is fully unsupervised, i.e. no threshold needs to be chosen. We show that maximum likelihood estimation can be performed by means of the EM algorithm and that the model is quite flexible in fitting data from different data-generating processes. Simulation experiments and a real-data application to automobiles claims suggest that the approach is equivalent in terms of goodness-of-fit, but easier to estimate, with respect to two existing distributions with similar features. All the methods are implemented in the \texttt{R} package \texttt{lognGPD}, available on CRAN.
}


\maketitle
\bmhead{Acknowledgments} This paper has been presented at the 15-th Scientific Meeting of the Classification and Data Analysis Group (Napoli, September 8-10, 2025). We would like to thank the
participants for their suggestions. We would also like to thank three anonymous reviewers whose valuable comments have considerably improved an earlier version of the paper.

\newpage

\section{Introduction}

In many fields of application, the population distribution is skewed and heavy-tailed. Examples range from simple instances such as exam grades or points scored by each player in a basketball game, to more complex phenomena measured at micro- or macro-level. To name just a few, notable instances are non-life insurance claims \citep{klu19}, income \citep{Goe23}, city size \citep{dac19}, international trade flows \citep{bee11a}, firm size \citep{axte01}.

In some of these cases, it may be difficult to find a single probability model for all the observations, especially if one needs to fit the tail with a high degree of precision. In other words, a common issue is the difference between the statistical features of the body and of the tail of the population distribution. 
Typical illustrations are loss data in insurance analytics \citep{klu19} and risk measurement \citep{pan06,cru15}, where the bulk of the data is well described by some classical size distribution, but the tail is heavier. An appropriate model for the latter is in most cases a Generalized Pareto distribution (GPD), which is a flexible and well-grounded model for observations above some large cutoff; see, e.g., \cite{mcn15}.

If the investigator only focuses on the tail, it may be enough to fit a GPD to the exceedances and disregard the remaining data: to this aim, Extreme Value Theory provides all the necessary tools \citep{emb97}. On the other hand, if one needs to fit the whole dataset, it is possible to ``combine'' in different ways a classical and a Pareto-type distribution.
For example, a distribution with a lognormal body and a Pareto-type tail can be obtained mainly in two ways. \cite{scoll07} proposes to splice a right-truncated lognormal and a Pareto distribution, %, see \cite{scoll07} and %\cite{bee15a} (continuous),\cite{bee24a}.
whereas, in the lognormal-GPD case, \cite{fri02} develop a dynamic mixture that gives more weight to the GPD as the observations get large. % and reconsidered by \cite{bee23} and \cite{bee24b}.

Both distributions will be formally introduced in Section \ref{sec:back}, but to motivate our proposal we anticipate here some of their features and limitations. The first model. i.e. the composite lognormal-Pareto developed by \cite{scoll07}, assumes that the tail follows an exact Pareto distribution; moreover, it requires continuity and differentiability constraints. The lognormal-GPD dynamic mixture of \cite{fri02} has an intractable likelihood, which complicates estimation. In both cases, despite the mixture representation, the EM algorithm cannot be used, essentially 
%The density of a dynamic mixture model is given by
%\begin{equation}
%	f(x;\boldsymbol{\theta})=\frac{p(x;\boldsymbol{\gamma}_1)f_1(x;\boldsymbol{\gamma}_2)+(1-p(x;\boldsymbol{\gamma}_1))f_2(x;\boldsymbol{\gamma}_3)}{Z},\quad x\in\mathbb{R}^+,
%	\label{eq:dens0}
%\end{equation}
%where $Z$ is a normalizing constant and $\boldsymbol{\gamma}_i$, $i=1,2,3$, are the parameter vectors of the weights $p(x;\boldsymbol{\gamma}_1)$, the body $f_1(x;\boldsymbol{\gamma}_2)$ and the tail $f_2(x;\boldsymbol{\gamma}_3)$, respectively. For the weight function it is convenient to employ the cumulative distribution function (cdf) of some continuous random variable, whereas $f_1$ and $f_2$ are continuous densities with positive support. \cite{fri02} use the Cauchy cdf, whereas \cite{bee24b} proposes the exponential cdf.
%Unfortunately, estimation of (\ref{eq:dens0}) is rather difficult. First, the normalizing constant cannot be written in closed form; second, the EM algorithm cannot be used,
because the components of the distribution (mixing weights and component densities) share all the parameters. As pointed out by \cite{fri02}, in the dynamic mixture case this remains an issue for any (non-constant) weight function. 

To the best of our knowledge, a static (i.e., with constant weights) mixture has never been considered. Its density is given by
\begin{equation}
f(x;\boldsymbol{\theta})=pf_1(x;\boldsymbol{\gamma}_1)+(1-p)f_2(x;\boldsymbol{\gamma}_2),\quad x\in\mathbb{R}^+,
\label{eq:static}
\end{equation}
where $p\in(0,1)$, and $f_1$ and $f_2$ are densities of continuous random variables with positive support. By definition, $\int_0^{\infty}f(x;\boldsymbol{\theta})dx=1$, so that there is no normalization issue. Maximum likelihood estimation (MLE) can be performed by means of the EM algorithm, since each component distribution has different parameters. Moreover, as long as both components are supported on $[0,+\infty)$, there is no threshold, and the density is automatically continuous and differentiable. In other words, the model inherits from the dynamic mixture setup its fully unsupervised nature, and does not require continuity and differentiability constraints.

In the following, we focus on the special case where $f_1$ and $f_2$ are the pdf of a $\text{Logn}(\mu,\sigma^2)$ and of a $\text{GPD}(0,\xi,\beta)$ random variable, respectively:
\begin{equation}
f(x;\boldsymbol{\theta})=pf_1(x;\mu,\sigma^2)+(1-p)f_2(x;0,\xi,\beta),\quad x\in\mathbb{R}^+,
	\label{eq:lognGPD}
\end{equation}
so that $\boldsymbol{\theta}=(p,\mu,\sigma^2,\xi,\beta)\in\boldsymbol{\Theta}\stackrel{\text{def}}{=}(0,1)\times\mathbb{R}\times\mathbb{R}^+\times\mathbb{R}\times\mathbb{R}^+$. It may be worth pointing out that, if $\xi$ is negative, the density is not differentiable at $-\beta/\xi$. However, $\xi$ is unlikely to be negative when modeling loss data.

Even though the M-step for the GPD parameters is not in closed form, fitting (\ref{eq:lognGPD}) is easier and computationally much lighter than estimating a dynamic mixture. 
In light of the requirements on the support of the two components and the skewed and leptokurtic features required for the resulting distribution, a size distribution different from the lognormal can be used without significantly changing the estimation methodology and the modeling capabilities. On the other hand, the zero-mean GPD seems to be the best choice, both for its theoretical properties as a tail model and because it is supported on $\mathbb{R}^+$.

\begin{figure}
	\begin{center}
		\includegraphics[width=12cm]{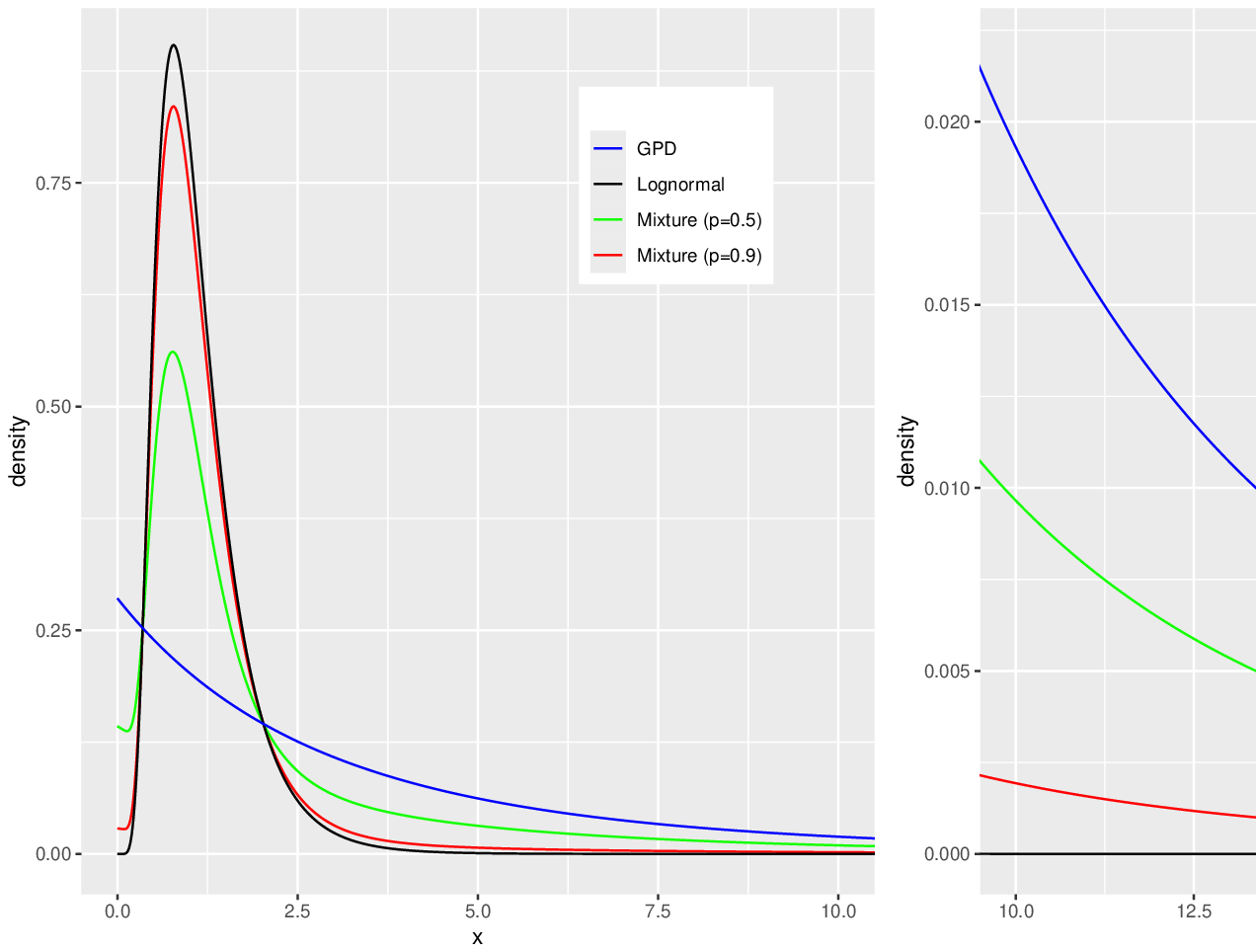} % figBase_0_0.5_0.25_3.5}
	\caption{Densities of static lognormal-GPD mixtures with $p\in\{0.9,0.5\}$, $\mu=0$, $\sigma^2=0.25$, $\xi=0.25$ and $\beta=3.5$, as well as of the lognormal and GPD component densities. The left panel shows the body ($x\le 10$), the right panel displays the tail ($x>10$).}% Lower panel: zoom on the distribution body.}
\label{fig:figbase}
% figBase.R
\end{center}
\end{figure}

Figure \ref{fig:figbase} displays the density (\ref{eq:lognGPD}) when $p\in\{0.9,0.5\}$, $\mu=0$, $\sigma^2=0.25$, $\xi=0.25$ and $\beta=3.5$, as well as the two component densities.
From the figure, we notice a few features. First, the GPD alone is unlikely to be a good model for the bulk of the data: this is not surprising, because the GPD with $\xi>0$ typically fits well only the  observations above a high threshold (see, e.g., \cite{emb97}). Second, the GPD tail is heavier than the lognormal one, hence the GPD contribution is key to model the largest observations. Third, the mixture combines the good fit of the lognormal to the body and of the GPD to the extreme observations, and the numerical value of $p$ allows one to obtain a distribution with a specific degree of tail heaviness. In Sect. \ref{sec:emp} we will actually be able to verify empirically that the observations in the body of the distributions are more likely to be lognormal, which confirms that the body is mostly described by the lognormal (see Fig. \ref{fig:app2}). Finally, it should be noticed that, if $\xi>-1$, the density is always bimodal, since it has a small mode at the origin. As can be seen in Fig. \ref{fig:figbase}, the mode is hardly noticeable; moreover, from the modeling point of view, when one's goal is to fit loss data and compute risk, this is usually a negligible problem, since small losses are seldom important.

In the present paper we investigate two main issues. The first one is estimation: the implementation and performance of the EM algorithm need to be assessed. The second one is flexibility: is the proposed model sufficiently flexible for data from different skewed and leptokurtic data-generating processes? % The main purpose of this paper is to tackle these issues.

The novelty of the paper is twofold. We propose a new model that combines a classical size distribution and a Pareto-type tail, yet remains unsupervised. This implies that, unlike most proposals in the literature, there is no threshold between the two distributions, whose estimation is typically difficult. In a second step we develop an EM algorithm for MLE, which is computationally cheaper than the computer-intensive approaches often necessary in similar setups. All in all, the model is flexible and easy to interpret, while the estimation procedure is reliable and reasonably fast.

The rest of the paper is organized as follows. In Section \ref{sec:back} we give some background about the problem of modeling non-negative data; in Section \ref{sec:lognGPD} we introduce the static lognormal-GPD mixture and detail the EM algorithm; in Section \ref{sec:sim} we illustrate the outcomes of simulation experiments, both in correctly- and mis-specified setups, and discuss some computational issues; in Section \ref{sec:emp} we fit the model to automobile claims data. Finally, in Section \ref{sec:concl} we discuss the results and conclude.

\section{Background}
\label{sec:back}

The literature about models that ``combine'' different distributions for the body and the tail is rather rich; their development has been motivated by actuarial applications (see, e.g., \citealp{klu19}) and by quantitative risk management problems \citep{pan06,cru15}.

A popular way of proceeding is based on splicing. A two-component spliced distribution is defined by the following density \citep{pan06}:
$$
f(x)=
\begin{cases}
a_1f_1(x), & c_0<x<c_1,\\
a_2f_2(x), & c_1<x<c_2,
\end{cases}
$$
where $a_1,a_2>0$, $a_1+a_2=1$, $f_1, f_2$ are density functions, and the definition can be extended in an obvious way to a finite number of components $k>2$. The family includes various different models, depending not only on the distributions used for the body and the tail, but also on the requirements about the junction point: the resulting density may be discontinuous, continuous or continuous and differentiable. The last two cases require constraints that reduce the number of free parameters.

Second, a closely related model is the dynamic mixture by \citet{fri02}% and reconsidered by \cite{bee23} and \cite{bee24b}
, which has the advantage of a smooth density with no need of finding a threshold, but is characterized by more complicated inferential procedures.

A special version of a spliced model that is also unsupervised, i.e. automatically determines the threshold, has been recently developed in a series of papers (see \citealp{dac23}, and the references therein) and is mainly aimed at cyber risk modeling. The idea is to join a lognormal and a GPD via a ``bridge'' distribution, which is chosen to be an exponential distribution. Continuity and differentiability constraints are used, so that the total number of free parameters is four, and estimation is carried out by means of an iterative algorithm.

More generally, our approach is related to other classes of models. First, in the actuarial setup, other mixture and spliced distributions have been proposed for heavy-tailed and censored or truncated losses: see, e.g., \cite{rey17}, \cite{blo19}, and \cite{bae24}. In a different field of applications, i.e. international trade, non-negative heavy-tailed commodities observations have also been modeled by means of flexible distributions such as the Tweedie distribution \citep{bar16}.

Second, the current work shares some features with an already established class of contamination-based approaches, based on mixtures of two components: one capturing the ``good'' observations and the other accounting for the ``bad'' ones; see, e.g., \cite{pun18} and \cite{pun19}. Finally, some works mainly focus on classification and clustering in a multi-group setup (e.g., \citealp{ban93}, \citealp{fra02}, \citealp{zha10}, \citealp{bro11}, \citealp{bro11}, \citealp{mcla16}, \citealp{mor19}). If we employ such models with only one group, they may work similarly to the static mixture proposed in the current paper.

Thus, one may consider a broader simulation-based and/or empirical comparison with other approaches, such as the contamination-based models mentioned above. Even though, to the best of our knowledge, a ``horse race'' analysis of this kind has never been carried out, we will not perform it here. Besides the practical reason that including it would require too much space in the current paper, we believe that these classes of models serve somewhat different purposes.

% there are some theoretical reasons that allow us to identify classes of models with different goals.}

Our approach is closer to the first class of models, which are mostly used for loss data, in particular in financial risk management and actuarial applications. Such data have two distinctive features: they are non-negative, so that the distributions must be supported on the positive real line, and heavy-tailed. Related to the latter feature, a precise estimate of the tail is usually crucial, which is why Pareto-type models are employed for the tail. This is the main reason why we only compare our approach to other distributions with Pareto-type tails. Overall, goodness-of-fit is more important than classification.

On the other hand, the second class of approaches, namely contaminated models, are mostly related not to loss data, but to real-valued datasets possibly containing a few outlying observations, or belonging to different groups. The main goal is usually a proper evaluation of the impact of such outliers on the bulk of the data, or estimation of clusters in the data. Unlike the previous case, tail behavior is not a genuine feature of the data-generating process, but results from contamination, so that tail estimation is typically not the main focus, whereas robust clustering/classification is essential.

Accordingly, in the rest of the paper, we will compare (\ref{eq:lognGPD}) to two specific cases, namely the composite lognormal-Pareto distribution \citep{scoll07}, termed Case 1 in the following, and the Cauchy-lognormal-GPD dynamic mixture \citep{fri02}, called Case 2 from now on. In particular, for these three distributions we will consider modeling capabilities, statistical properties of parameter estimation procedures, and computational costs.

\subsection{Case 1: the composite lognormal-Pareto distribution}

The continuous and differentiable composite lognormal-Pareto density \citep{scoll07} is given by
\begin{equation}
	\label{eq:mix1}
	f(x)=r(\sigma^2,\alpha) f_1(x;\mu(x_{min},\alpha,\sigma^2),\sigma^2,x_{min})+(1-r(\sigma^2,\alpha)) f_2(x;x_{min},\alpha), %\quad x>0,
\end{equation}
where
\begin{equation}
	\label{eq:tr_logn}
	f_1(x;\mu,\sigma^2,x_{min})=\frac{1}{\Phi\left(\frac{\log(x_{min})-\mu}{\sigma}\right)}\frac{1}{x\sigma\sqrt{2\pi}}e^{-\frac{1}{2}\left(\frac{\log(x)-\mu}{\sigma}\right)^2}\mathbb{I}_{\{0< x\le x_{min}\}}
\end{equation}
is the $\text{Logn}(\mu,\sigma^2)$ density right-truncated at $x_{min}$ and $f_2(x;x_{min},\alpha)$ is the Pareto density with scale and shape parameters $x_{min}$ and $\alpha$, respectively. From the definition of the density, we see that this is a spliced distribution. Due to the continuity and differentiability constraints, $r(\sigma^2,\alpha)$ and $\mu(x_{min},\alpha,\sigma^2)$ are given by
\begin{align}
r(\sigma^2,\alpha)&=\frac{\sqrt{2\pi}\alpha\sigma\Phi(\alpha\sigma)e^{\frac{1}{2}(\alpha\sigma)^2}}{\sqrt{2\pi}\alpha\sigma\Phi(\alpha\sigma)e^{\frac{1}{2}(\alpha\sigma)^2}+1},\nonumber \\
\mu(x_{min},\alpha,\sigma^2)&=\log(x_{min})-\alpha\sigma^2.\nonumber
\end{align}
Therefore, only $\sigma^2$, $\alpha$ and $x_{min}$ are free parameters.%, whereas $\mu$ is a function of $\sigma^2$, $\alpha$ and $\theta$, and $r$ is a function of $\sigma^2$ and $\alpha$.

\subsection{Case 2: the Cauchy-lognormal-GPD dynamic mixture}
\label{sec:Case2}

The density is given by
\begin{equation}
\label{eq:fridens}
f(x;\boldsymbol{\theta})=\frac{(1-p(x;\mu_c,\tau))f_1(x;\mu,\sigma^2)+p(x;\mu_c,\tau)f_2(x;\beta,\xi)}{Z},
\end{equation}
where $x\in\mathbb{R}^+$, $\mu_c,\mu\in\mathbb{R}$, $\tau,\sigma^2,\xi,\beta\in\mathbb{R}^+$, $\boldsymbol{\theta}=(\mu_c,\tau,\mu,\sigma^2,\beta,\xi)'$, $f_1(x;\mu,\sigma^2)$ is the lognormal density with parameters $\mu$ and $\sigma^2$, $f_2(x;\beta,\xi)$ is the zero-location GPD density with scale and shape parameters equal to $\beta$ and $\xi$, respectively. The normalizing constant $Z$ cannot be computed in closed form: it is equal to (for details see \citealp{bee23})
$$
Z=Z(\boldsymbol{\theta})=1+\frac{1}{\pi}I,
$$
where
$$
I=\bigintsss_0^\infty\left[\frac{1}{\beta}\left(1+\frac{\xi x}{\beta}\right)^{-1/\xi-1}-\frac{1}{\sqrt{2\pi}\sigma x}e^{-\frac{1}{2}\left(\frac{\log{x}-\mu}{\sigma}\right)^2}\right]\arctan\left(\frac{x-\mu_c}{\tau}\right)dx.
$$
The original version of this model, with the lognormal density replaced by the Weibull, had been developed by \cite{fri02} and was also characterized by the normalizing constant issue.
%\medskip

\section{A static lognormal-GPD mixture}
\label{sec:lognGPD}

In the present paper we implement the EM algorithm for MLE of the finite mixture with density (\ref{eq:static}). In the following we will call it a \emph{static mixture}, to avoid confusion with the dynamic mixtures of Section \ref{sec:Case2}. As usual in finite mixture analysis (e.g., \citealp{mcl00}), MLE would be easy if population membership were known, and we are now going to exploit this feature to implement the EM algorithm.
%Since complete-data MLE of each population is rather straight

\subsection{Maximum Likelihood Estimation: the EM algorithm}
\label{sec:EM}

Before tackling estimation, we briefly comment on identifiability of the static lognormal-GPD mixture. To avoid trivialities, assume $p\neq 0$ and $p\neq1$. In this case, since the lognormal is never identical to the GPD for any numerical values of the parameters, the mixture should be identifiable.

Let $\boldsymbol{x}=(x_1,\dots,x_n)$ be the observed data, i.e., a random sample from a mixture distribution. Let $\boldsymbol{z}=(z_{1j},\dots,z_{nj})$ represent the missing data, with $z_{ij}$ ($i=1,\dots,n$) being an indicator variable of population membership: $z_{ij}=1$ if the $i$-th observation belongs to the $j$-th population, and $z_{ij}=0$ otherwise.
Let $\ell(\boldsymbol{\theta})$ be the observed log-likelihood function obtained from the mixture density, with $\boldsymbol{\theta}\in\boldsymbol{\Theta}\subset\mathbb{R}^d$ a vector of parameters.
Let $\boldsymbol{v}=(\boldsymbol{x}',\boldsymbol{z}')'$ be the complete data, with density and complete log-likelihood denoted by $g_c(\boldsymbol{v};\boldsymbol{\theta})$ and $\ell_c(\boldsymbol{\theta})$ respectively.
At the $t$-th iteration, the algorithm is carried out by means of two steps.

\medskip
\noindent\textbf{E-step}. Compute the conditional expectation of $\ell_c(\boldsymbol{\theta})$, given the current value of $\boldsymbol{\theta}$ and the observed sample
$\boldsymbol{x}$:%. Conditionally on the vector of current parameter values $\boldsymbol{\theta}^{(t)}$, the E-step is equal to
\begin{equation}
	\label{eq:Estep0}
	Q(\boldsymbol{\theta};\boldsymbol{\theta}^{(t)})\stackrel{\text{def}}{=}\text{E}_{\boldsymbol{\theta}^{(t)}}\{\ell_c(\boldsymbol{\theta})|\boldsymbol{x}\}.
\end{equation}
%\medskip
\textbf{M-step}. Maximize, with respect to $\boldsymbol{\theta}$, the so-called $Q$-function, i.e. the conditional expectation of $\ell_c(\boldsymbol{\theta})$ provided by the E-step (\ref{eq:Estep0}): %, so that $\boldsymbol{\theta}^{(t+1)}$ is given by
\begin{equation}
	\label{eq:Mstep}
	\boldsymbol{\theta}^{(t+1)}=\argmax_{\boldsymbol{\theta}} Q(\boldsymbol{\theta};\boldsymbol{\theta}^{(t)}).
\end{equation}

\medskip
The E- and M-step (\ref{eq:Estep0}) and (\ref{eq:Mstep}) are then iterated until convergence is reached according to some stopping rule.
The algorithm monotonically increases the observed likelihood at each iteration; under regularity conditions (\citealp{wu83}), the sequence $\boldsymbol{\theta}^{(t)}$ converges to a stationary point and the estimators are asymptotically efficient.

Let us now turn to the present setup: let $\boldsymbol{x}=(x_1,\dots,x_n)$ be the observed data, i.e., a random sample from the static lognormal-GPD mixture (\ref{eq:lognGPD}). Let $\boldsymbol{z}=(z_1,\dots,z_n)$ represent the missing data, with $z_i=1$ if the $i$-th observation is lognormal, and $z_i=0$ if it is GPD.
The observed and complete log-likelihood functions can be written as:
\begin{align}
\ell(\boldsymbol{\theta};\boldsymbol{x})&=\sum_{i=1}^n\log\{pf_1(x;_i\mu,\sigma^2)+(1-p)f_2(x_i;\xi,\beta)\},\nonumber \\
\ell_c(\boldsymbol{\theta};\boldsymbol{x},\boldsymbol{z})&=\sum_{i=1}^nz_i\log pf_1(x_i;\mu,\sigma^2)+\sum_{i=1}^n(1-z_i)\log (1-p)f_2(x_i;\xi,\beta).\nonumber
\end{align}
The conditional expectation of the complete log-likelihood function is given by
\begin{equation}
\text{E}(\ell_c(\boldsymbol{\theta};\boldsymbol{x})|\boldsymbol{x},\boldsymbol{\theta}^{(t)})=\sum_{i=1}^n\tau_{i1}^{(t)}\log f_1(x;_i\mu^{(t)},\sigma^{2(t)})+\sum_{i=1}^n\tau_{2i}^{(t)}\log f_2(x_i;\xi^{(t)},\beta^{(t)}),
\label{eq:Qfunc}
\end{equation}
$\tau_{2i}^{(t)}=1-\tau_{i1}^{(t)}$ and
\begin{equation}
\tau_{i1}^{(t)}=\frac{p^{(t)}f_1(x_i;\mu^{(t)},\sigma^{2(t)})}{p^{(t)}f_1(x_i;\mu^{(t)},\sigma^{2(t)})+(1-p^{(t)})f_2(x_i;\xi^{(t)},\beta^{(t)})}.
\label{eq:Estep}
\end{equation}
Equation (\ref{eq:Estep}), which is the E-step of the algorithm, represents the posterior probability, at the $t$-th iteration, that the $i$-th observation belongs to the lognormal population. At convergence, this quantity can be used for classification purposes.

As for the M-step, it results from the maximization of (\ref{eq:Qfunc}) with respect to all parameters. The M-steps for $p$, $\mu$ and $\sigma^2$ are given in closed form by 
\begin{align}
p^{(t)}&=\frac{1}{n}\sum_{i=1}^n\tau_{i1}^{(t)};\nonumber \\
\mu^{(t)}&=\frac{1}{np^{(t)}}\sum_{i=1}^n\tau_{i1}^{(t)}\log x_i;\nonumber \\
\sigma^{2(t)}&=\frac{1}{np^{(t)}}\sum_{i=1}^n\tau_{i1}^{(t)}(\log x_i-\mu^{(t)})^2.\nonumber
\end{align}
On the other hand, even in the complete-data case, the MLEs of $\xi$ and $\beta$ can only be found via numerical maximization. Hence, the M-steps for the GPD parameters are not explicit. Specifically, they are the solution of the maximization of the second summand of (\ref{eq:Qfunc}):
\begin{equation}
\xi^{(t)},\beta^{(t)}=\arg\max_{\xi,\beta}\sum_{i=1}^n\tau_{2i}\log f_2(x_i;\xi,\beta).
\label{eq:MGPD}
\end{equation}
The function $\sum_{i=1}^n\tau_{2i}\log f_2(x_i;\xi,\beta)$ is essentially a weighted version of the GPD log-likelihood function and can be maximized analogously, by means of standard optimization routines.

Practical implementation of the algorithm requires specification of initial values $p^{(0)}$, $\mu^{(0)}$, $\sigma^{2(0)}$, $\xi^{(0)}$ and $\beta^{(0)}$. For the mixing weight we set
$$
p^{(0)}=\frac{\#\{y:y<\text{median}(y)\}}{n}.%\ \mu^{(0)}=\frac{1}{n}\sum_{i=1}^n \log y_i,\ \sigma^{2(0)}=\frac{1}{n}\sum_{i=1}^n (\log y_i-\mu^{(0)})^2,
$$
For $\mu^{(0)}$ and $\sigma^{2(0)}$ we use the lognormal MLEs, whereas $\xi^{(0)}$ and $\beta^{(0)}$ are given by the GPD MLEs, in both cases computed with all the observations. 

The initialization of the EM algorithm is typically important when the log-likelihood function is multimodal, because in this case convergence of an EM sequence to a global or local maximizer, or to a saddle point, depends on the starting point \citep[Sect. 3.4]{mcl08}. Consequently, we have double-checked the impact of different initializations: even significant changes in initializations (for example, $p_0=0.2$ $p_0=0.8$ $\mu_0$ and $\sigma_0$ equal to the lognormal MLEs computed with the observations below the median, $\xi_0$ and $\beta_0$ computed with the observations above the median) result in small changes of the results, mostly when $n=100$. %We believe that this is one of the strengths of this approach, since finding initial values is typically a problem also in competing models.}

For the EM algorithm, stopping rules are usually based on subsequent values of either estimated parameters in the M-step, or of maximized log-likelihood: e.g., \citet[Sect. 4.9]{mcl08}, or \citet[p. 661]{flu97}. In this paper we use the first approach; in particular, the algorithm stops when $\max_{1\le i\le 5}|\theta_i^{(t)}-\theta_i^{(t-1)}|<\epsilon$, with $\boldsymbol{\theta}=(p,\mu,\sigma^2,\xi,\beta)'$ and $\epsilon=10^{-6}$. In some of the simulations we have also tried monitoring successive log-likelihood values, and the outcomes are essentially identical.

\section{Simulation experiments}
\label{sec:sim}

The experiments in this section have a twofold goal. First, we aim at assessing the finite-sample behavior of the estimators in the correctly-specified setup, that is, when data are simulated from (\ref{eq:lognGPD}); this will be called ``Correctly-specified setup'' in the following. Second, we sample from the data-generating processes (DGPs) of Case 1 (Section \ref{sec:case1}) and Case 2 (Section \ref{sec:case2}) and compare the static mixture with the two models (\ref{eq:mix1}) and (\ref{eq:fridens}). In other words, we estimate our model in two different mis-specified setups, to study the properties of parameter estimators; see Sect. \ref{sec:case1} and \ref{sec:case2} below. Finally, we also perform a small experiment where all models are mis-specified: in particular, we will fit the distributions to data simulated form using data from the Generalized Beta Distribution of the second kind.

In all setups, we also estimate the Value-at-Risk (VaR), which is given by the smallest number $\ell$ such that the probability that the loss on an asset or portfolio $L$ exceeds $\ell$ is no larger than $(1-\alpha)$, where $\alpha\in(0,1)$ is a predetermined confidence level \citep{mcn15}.

\subsection{Correctly-specified setup}

\begin{figure}[!h]
	\begin{center}
		\includegraphics[width=\textwidth,angle=0]{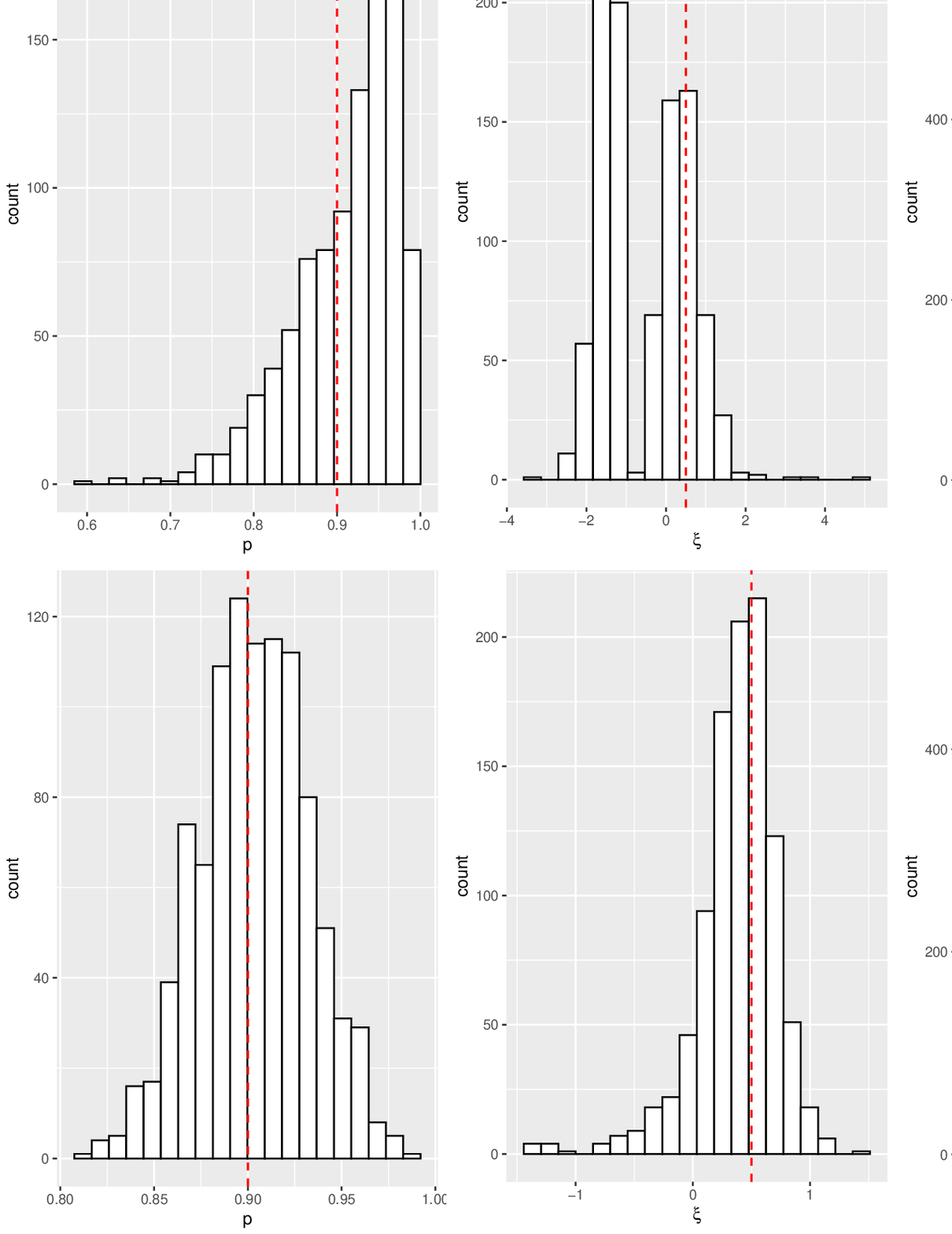}
		\caption{Distributions of the parameter estimates for $n=100$ (first row) and $n=500$ (second row) in the correctly-specified setup, when $\xi=0.5$. The red vertical lines represent the true values of the parameters.}
	\label{fig:hist}
	% ResultsSimMiss.R
\end{center}
\end{figure}

We simulate observations from (\ref{eq:lognGPD}) with $p=0.9$, $\mu=0$, $\sigma^2=0.25$, $\xi\in\{0.25,0.5\}$, $\beta=3.5$, for sample sizes $n\in\{100,500\}$. When $\xi=0.5$, the distributions of the EM-based estimates $\hat{p}$, $\hat{\xi}$ and $\hat{\beta}$ are shown in Figure \ref{fig:hist}\footnote{the distributions of the estimates of $\mu$ and $\sigma$ are quite ``well-behaved'' for both sample sizes. Hence, we omit them to save space.}. The number of replications is $B=1000$. As can be seen, the properties of the estimators considerably improve when $n=500$: the variance is smaller, there are fewer outliers and the distributions of $\hat{p}$ and $\hat{\xi}$ are more ``bell-shaped''.

%\FloatBarrier

Table \ref{tab:bias} shows the bias, median-bias, root-mean-squared-error (RMSE) and median-RMSE (RMSE$_{med}$) of the estimators. Letting $\tilde{\theta}_i=\text{median}_{j=1,\dots,B}\theta_{i,j}$ ($i=1,\dots,d$), where $\theta_{i,j}$ are the $B$ simulated values of the $i$-th parameter, the median-RMSE is given by
$$
RMSE_{med}(\hat{\theta}_i)=\sqrt{b_{med}(\theta_i)^2 + mad(\theta_i)^2},
$$
where $b_{med}(\theta_i)\stackrel{\text{def}}{=}\tilde{\theta}_i-\theta_i$ and $mad(\theta_i)\stackrel{\text{def}}{=}\text{median}_{j=1,\dots,B}|\theta_{i,j}-\tilde{\theta}_i|$ are the median-bias and the median absolute deviation, respectively. The reason why we also compute median-based accuracy measures is their robustness to outliers. In all tables, the true values of $\mu$ and $\sigma$ are $\mu=0$ and $\sigma=0.5$.

\begin{table}[htbp]
	\centering
	\caption{Bias, median-bias, RMSE and RMSE$_{med}$ when $p=0.9$, $\xi=0.5$ and $\beta=3.5$.}
%	\bigskip
	\begin{tabular}{c|cccccc}
		\hline
		&  & $p$  & $\mu$  & $\sigma$ & $\xi$  & $\beta$ \\
		\hline
		$n=100$ & bias & 0.013 & 0.005 &  0.000 & $-1.053$ & 14.515 \\
		& b$_{med}$	& 0.030 & 0.004 & 0.002 & $-1.563$ & 5.611 \\
\hline
		$n=500$ & bias & $-0.003$ & $-0.002$ & $-0.000$ & 0.127 & $-1.390$ \\
		& b$_{med}$	& 0.004 &  0.002 &  0.000 & $-0.087$ & 0.350 \\
		\midrule
	$n=100$ & RMSE & 0.063 & 0.060 &  0.054 &  1.509 & 32.702 \\
	& RMSE$_{med}$	& 0.061 &  0.060 &  0.051 &  2.008 & 11.788 \\
	\hline
	$n=500$ & RMSE & 0.001 &  0.001 &  0.001 &  0.133 & 16.801 \\
	& RMSE$_{med}$	& 0.030 & 0.026 & 0.024 & 0.287 & 1.675 \\
	\midrule
	\end{tabular}
	\label{tab:bias}
\end{table}

For $n=500$, the median-based bias and RMSE of $\hat{\mu}$ and $\hat{\sigma}$ show no obvious changes with respect to the corresponding mean-based measures, whereas for $\hat{\xi}$, and especially $\hat{\beta}$, the measures are different. When $n=100$ there are considerable differences, in line with the outliers noted in Figure \ref{fig:hist}. This suggests that, for small sample size, and also for $n=500$ as long as we are concerned with the GPD parameters, the algorithm sometimes does not converge to a global maximum. Notice that this issue mostly affects the GPD parameters: since the outcomes refer to the $p=0.9$ setup and consequently, when $n=100$, the number of GPD observations is very small, this is not surprising. In particular, the histogram of $\hat{\xi}$ is roughly bimodal, with the left-hand mode quite far away from the true value. To get some intuitive insight into this issue, we double-checked the results of the replications, and found that the negative values of $\hat{\xi}$ usually correspond to values of $\hat{\pi}$ close to 1. If this is the case, $\tau_{2i}$ is mostly close to zero, and estimating the GPD parameters becomes more difficult.

\subsection{Mis-specified models}

The outcomes of the previous section suggest that the EM-based fitting procedure of the static lognormal-GPD mixture is effective in a correctly-specified setup, unless the sample size is small. However, in general, the algorithm may suffer from instability issues. Hence, to check whether the model is suitable also for other DGPs, in this section we study its goodness of fit when the true distribution is different from (\ref{eq:lognGPD}). In particular, we fit the model to data from the composite lognormal-Pareto distribution developed by \cite{scoll07}, termed Case 1 in Section \ref{sec:back}, and the Cauchy-lognormal-GPD dynamic mixture proposed by \cite{fri02}, called Case 2 in Section \ref{sec:back}. 

\subsubsection{Case 1: the composite lognormal-Pareto distribution}
\label{sec:case1}

\begin{figure}[!h]
	\begin{center}
		\includegraphics[width=11cm,angle=0]{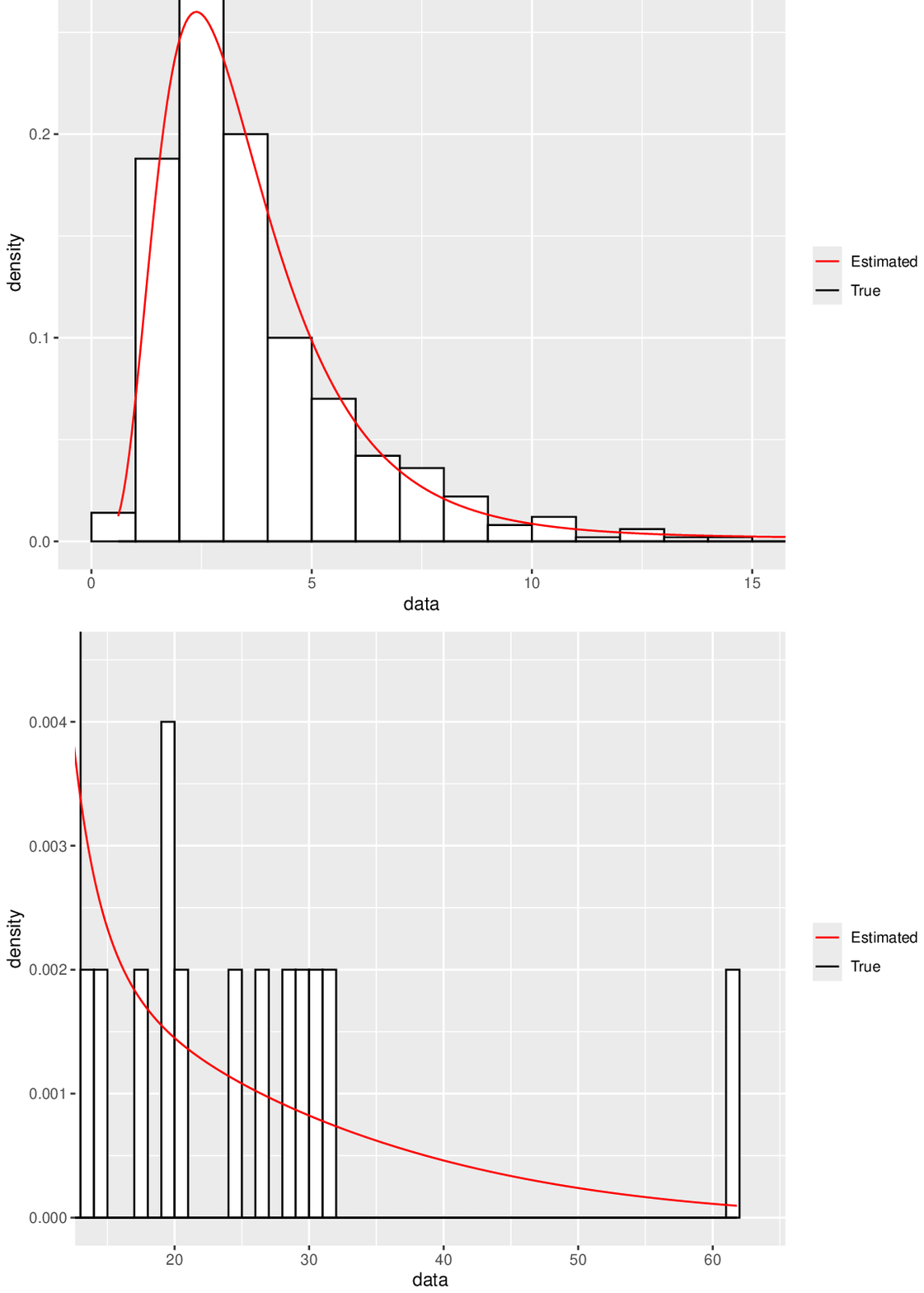}
		\caption{Histogram of a simulated dataset from Case 1 ($n=500$, $\mu = 0$, $\sigma = .5$, $x_{min} = 5$ and $\alpha = 2$), along with the true composite lognormal-Pareto density and the estimated lognormal-GPD static mixture density. Upper panel: distribution body ($x<15$). Lower panel: distribution tail ($x>15$). The bin width is the same in both panels.}
		\label{fig:miss2}
	\end{center}
\end{figure}

Figure \ref{fig:miss2} displays the histogram of 500 observations sampled from the composite distribution of Case 1, with parameters $\mu = 0$, $\sigma = 0.5$, $x_{min} = 5$ and $\alpha = 2$. The true density and the estimated static mixture with parameters equal to the mean of the 1000 estimated parameter vectors obtained in the simulation experiments are displayed as well.

Table \ref{tab:miss4} shows estimates and standard errors of the static mixture parameters when the shape parameter of the true DGP, i.e. the composite lognormal-Pareto, is $\alpha\in\{1.5,2\}$; in addition, we report the \textit{p}-values of the Kolmogorov-Smirnov (KS) and Anderson-Darling (AD) goodness-of-fit (GoF) tests. The null hypothesis of both tests is that two samples with the same sample size arise from a common unspecified distribution function. KS is implemented via the \texttt{ks.test} function, AD by means of the \texttt{ad.test} from the \texttt{kSamples} package \citep{CRAN:kSamples}. 
% ...\R2\R\Fig3_R2.R
%\FloatBarrier
\begin{table}[htbp]
	\centering
	\caption{Case 1: parameter estimates, standard errors and \textit{p}-values of the GoF tests when the static mixture is fitted to data sampled from the composite lognormal-Pareto (Case 1) with $\mu = 0$, $\sigma = 0.5$, $x_{min} = 5$ and $\alpha\in\{1.5,2\}$.}
%	\bigskip
	\begin{tabular}{c|ccccccc}
		\hline
		& $p$  & $\mu$  & $\sigma$ & $\xi$  & $\beta$ & KS & AD \\
		\hline
		$n=100$, $\alpha=1.5$ & 0.895 &  1.327 &  0.563 & $-0.545$ & 87.531 & 0.638 & 0.620  \\ % when n=100, 2.8% of non-convergence
	& (0.082) &  (0.073) &  (0.079) & (0.815) & (780.324) & (0.272) & (0.259) \\
	\hline
	$n=500$, $\alpha=1.5$ & 0.875 &  1.307 &  0.549 &  0.194 & 19.285  & 0.551 & 0.494\\
	& (0.037) & (0.029) & (0.030) & (0.246) & (9.088) & (0.261) & (0.242) \\
\midrule
		$n=100$, $\alpha=2$ & 0.943 & 1.162 &  0.537 & $-0.918$ & 42.254 & 0.657 & 0.643 \\ % when n=100, 5.5% of non-convergence
			& (0.061) & (0.058) & (0.064) & (0.772) & (78.148) & (0.254) & (0.254) \\
\hline
		$n=500$, $\alpha=2$ & 0.933 &  1.147 &  0.528 & $-0.181$ & 20.046 & 0.593 & 0.564 \\
		& (0.028) & (0.025) & (0.026) & (0.441) & (18.203) & (0.265) & (0.251) \\
		\midrule
	\end{tabular}
	\label{tab:miss4}
\end{table}
As for parameter estimates, the standard deviation of $\hat{\beta}$ when $n=100$ is noticeably large. Nevertheless, both Figure \ref{fig:miss2} and the GoF tests in Table \ref{tab:miss4} suggest a more complex model is not needed.

Since these kinds of distributions are expected to be especially important for risk measurement purposes, Table \ref{tab:VaRmiss2} displays the estimated VaR obtained via the correctly-specified composite lognormal-Pareto model and the mis-specified lognormal-GPD static mixture, along with the true quantile. The VaR is computed via Monte Carlo simulation with $10^4$ replications.
95\% confidence intervals are reported as well. The point estimates of the VaRs are close to each other and, as clearly results from the confidence intervals, they are not significantly different from each other.

\begin{table}[htbp]
	\centering
	\caption{Case 1: estimated VaR when the static mixture is fitted to data sampled from the composite lognormal-Pareto (Case 1) when $\sigma = .5$, $\xi = 2$, $x_{min} = 5$; 95\% confidence intervals are in brackets. ``Correct-spec.'' refers to the VaR computed as a quantile of the estimated composite lognormal-Pareto distribution, ``Mis-spec.'' to the VaR computed as a quantile of the estimated static lognormal-GPD mixture.}
%	\bigskip
	\begin{tabular}{cc|ccc}
		\midrule
		&	& 0.95  & 0.99  & 0.995 \\
		\midrule
		$n=100$ & Correct-spec. & 10.953 & 25.378 & 36.594 \\
		& & [8.349,14.534] & [13.688,43.223] & [16.970,68.417] \\
\midrule
		$n=100$ & Mis-spec. & 10.877 & 27.444 & 34.921 \\
		& & [8.144,17.683] & [12.155,73.428] & [13.408,114.102] \\
\midrule
		$n=500$ & Correct-spec. & 10.605 & 23.801 & 33.679 \\
		& & [9.291,11.986] & [18.104,30.524] & [23.954,44.842] \\
\midrule
		$n=500$ & Mis-spec. & 9.968 & 27.682 & 37.911 \\
		& & [8.664,12.228] & [16.303,42.868] & [20.867.62.658] \\
\midrule
		True &	& 10.568 & 23.630 & 33.346 \\
		\midrule
	\end{tabular}
	\label{tab:VaRmiss2}
	%	mainSimMisspecified1.R, resultsSimMiss.R, VaRwellSpec1.R, ResultsVaRWellSpec.R
\end{table}

\subsubsection{Case 2: the Cauchy-lognormal-GPD dynamic mixture}
\label{sec:case2}

\begin{figure}[!h]
	\begin{center}
		\includegraphics[width=11cm,angle=0]{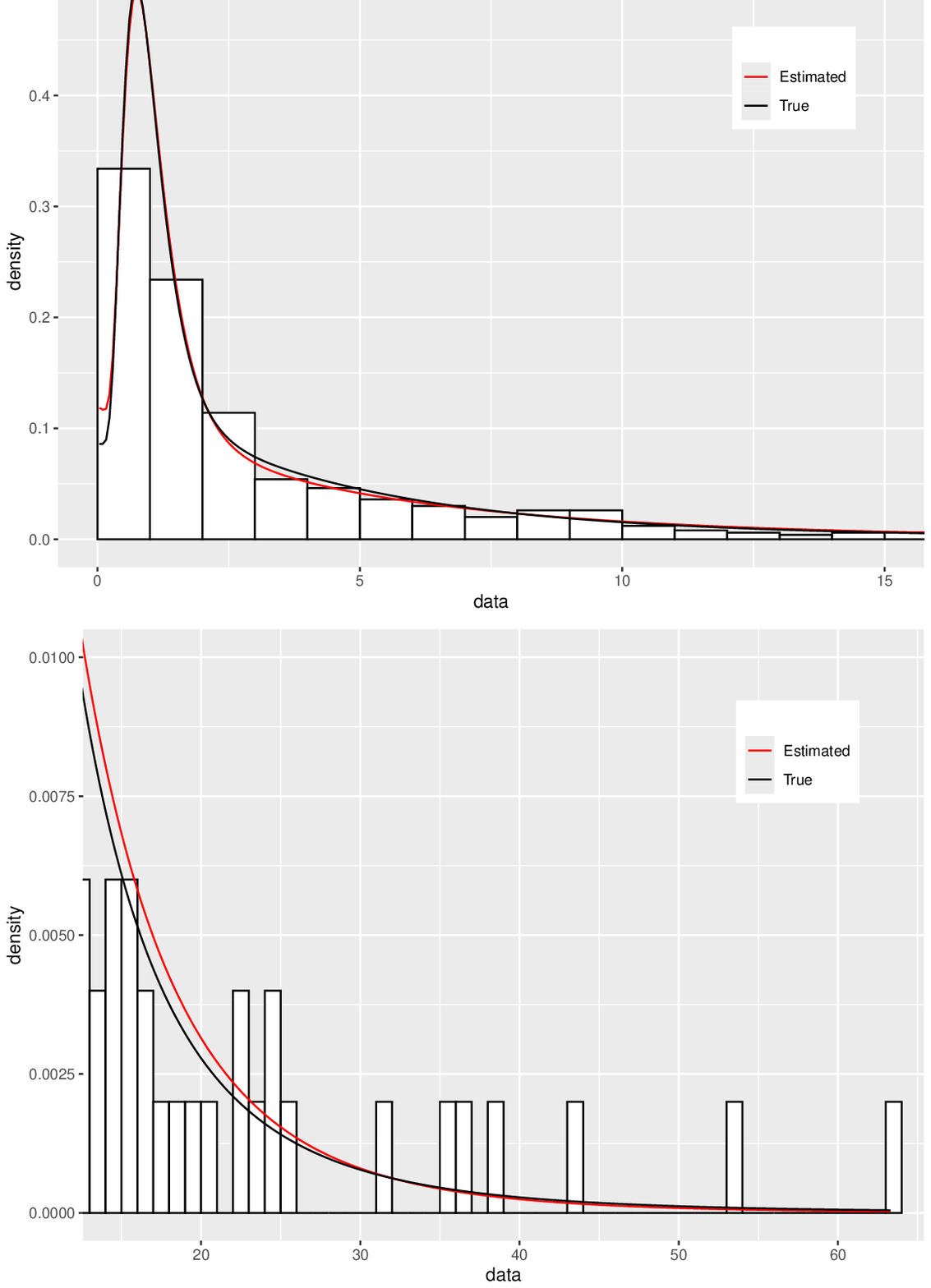}
\caption{Histogram of the simulated data from Case 2 ($n=500$, $\mu_c = 1$, $\tau = 2$, $\mu = 0$, $\sigma^2 =0.25$, $\xi = 0.25$ and $\beta = 3.5$) along with the true dynamic mixture density and the estimated lognormal-GPD static mixture density. Upper panel: distribution body ($x<15$). Lower panel: distribution tail ($x>15$). The bin width is the same in both panels.}
		\label{fig:miss1}
		% ResultsSimMiss.R
	\end{center}
\end{figure}

We perform the same analysis of Section \ref{sec:case1} when the true DGP is the Cauchy-lognornal-GPD distribution (Case 2 in Section \ref{sec:back}). Figure \ref{fig:miss1} displays the histogram of 500 observations sampled from the dynamic mixture (Case2) with parameters $\mu_c = 1$, $\tau = 2$, $\mu = 0$, $\sigma^2 =0.25$, $\xi = 0.25$ and $\beta = 3.5$. Superimposed to the histogram are the true dynamic mixture density and the estimated static mixture with parameters equal to the mean of the 1000 estimated parameter vectors obtained in the simulation experiments.

% histMiss_',muc,'_',tau,'_',mu,'_',sigma,'_',xi,'_',beta,'_',N,'_',nreps,'.eps'
\FloatBarrier

Table \ref{tab:miss2} displays parameter estimates, standard errors and $p$-values of the GoF tests when the static mixture is fitted to data sampled from the dynamic mixture (Case 2) with $\xi\in\{0.25,0.5\}$. The tests confirm that the static lognormal-GPD mixture fits quite well the data.

Table \ref{tab:VaRmiss} reports the correctly-specified VaR, computed using the dynamic mixture with parameters estimated via MLE, and the mis-specified VaR estimated from the static mixture, with corresponding standard errors.

\begin{table}[htbp]
	\centering
	\caption{Case 2: parameter estimates, standard errors and \textit{p}-values of the GoF tests when the static mixture is fitted to data sampled from the dynamic mixture (Case 2) with $\mu_c = 1$, $\tau = 2$, $\mu = 0$, $\sigma^2 =0.25$, $\xi\in\{0.25,0.5\}$ and $\beta = 3.5$.}
%	\bigskip
	\begin{tabular}{c|ccccccc}
		\midrule
		& $\pi$  & $\mu$  & $\sigma$ & $\xi$  & $\beta$ & KS & AD \\
		\midrule
	$n=100$, $\xi=0.25$ & 0.442 & $-0.017$ & 0.488 & $-0.001$ & 7.188 & 0.690 & 0.673 \\ % when n=100, 1.1% of non-convergence
		& (0.152) & (0.176) & (0.165) & (0.363) & (9.278) & (0.253) & (0.253) \\
\midrule
		$n=500$, $\xi=0.25$ & 0.403 & $-0.048$ & 0.480 &  0.131 &  5.002 & 0.653 & 0.626\\
		& (0.053) & (0.063) & (0.060) & (0.075) & (0.603) & (0.264) & (0.272) \\
		\midrule
		$n=100$, $\xi=0.5$ & 0.434 & $-0.0188$ & 0.492 &  0.275 &  7.963 & 0.675 & 0.674  \\ % when n=100, 1.1% of non-convergence
		& (0.140) &  (0.178) & (0.163) & (0.350) & (15.937) & (0.255) & (0.249) \\
\midrule
		$n=500$, $\xi=0.5$ & 0.398 & $-0.051$ & 0.477 & 0.381 & 5.312 & 0.642 & 0.628\\
		& (0.050) & (0.060) & (0.058) & (0.087) & (0.713) & (0.257) & (0.258) \\
		\midrule
	\end{tabular}
	\label{tab:miss2}
\end{table}

\begin{table}[htbp]
	\centering
	\caption{Case 2: estimated VaR when the static mixture is fitted to data sampled from the dynamic mixture (Case 2) with $\mu_c = 1$, $\tau = 2$, $\mu = 0$, $\sigma^2 =0.25$, $\xi = 0.25$ and $\beta = 3.5$. 95\% confidence intervals are in brackets. ``Correct-spec.'' refers to the VaR computed as a quantile of the estimated dynamic lognormal-GPD mixture, ``Mis-spec.'' to the VaR computed as a quantile of the estimated static lognormal-GPD mixture.}
%	\bigskip
	\begin{tabular}{cc|ccc}
		\midrule
		&	& 0.95  & 0.99  & 0.995 \\
		\midrule
		$n=100$ & Correct-spec. & 14.218 & 29.056 & 38.241 \\
		& & [9.665,20.191] & [16.831,54.053] & [20.023,79.467] \\
\midrule
		$n=100$ & Mis-spec. & 14.549 & 26.604 & 32.613 \\
		& & [10.121,20.035] & [14.497,44.409] & [15.975,58.978] \\
\midrule
		$n=500$ & Correct-spec. & 14.262 & 28.694 & 37.028 \\
		& & [12.101,16.144] & [21.716,37.209] & [26.199,51.662] \\
\midrule
		$n=500$ & Mis-spec. & 14.602 & 27.146 & 33.479 \\
		& & [12.445,17.081] & [21.414,35.180] & [25.055.44.737] \\
\midrule
		True &	& 14.211 & 28.352 & 36.342 \\
		\midrule
	\end{tabular}
	\label{tab:VaRmiss}
	%	mainSimMisspecified1.R, resultsSimMiss.R, VaRwellSpec1.R, ResultsVaRWellSpec.R
\end{table}

\FloatBarrier

The estimated VaRs are again very close to each other, and the differences are non-significant, as shown by the large overlap of the confidence intervals. In this setup, unlike Case 1, the static mixture has narrower confidence intervals, presumably because the dynamic mixture has one parameter more, and overall its estimation is difficult.

\subsubsection{Case 3: the Generalized Beta Distribution of the Second Kind}
\label{sec:case3}

We finally consider a setup where all models considered so far are mis-specified. To this aim, we simulate observations from a Generalized Beta Distribution of the second kind (GB2; see \citealp[Sect. 6.1]{klei03}) with parameters $a=180$, $b=2$, $p=q=0.7$, where $b$ is a scale parameter, whereas $a$, $p$ and $q$ are shape parameters. We sample $n=500$ observations from the GB2(180, 2, 0.7, 0.7) distribution, fit the static mixture (\ref{eq:lognGPD}), the composite-lognormal-Pareto distribution (Case 1) and the Cauchy-lognormal-GPD dynamic mixture (Case 2), compute the KS and AD tests, as well as the VaR at levels 95, 99 and 99.5\%; to double-check the goodness-of-Fit, we also compute the Bayesian Information Criterion (BIC). This procedure is repeated $B=500$ times. The simulated observations are strongly right-skewed and heavy-tailed: the average skewness and kurtosis of the simulate data across the replications are $sk=9.33$ and $ku=122.25$, respectively. The outcomes are reported in Table \ref{tab:GB2}.

\begin{table}[htbp]
	\centering
	\caption{Case 3: estimated VaR and GoF tests when the three models are fitted to data sampled from a GB2 distribution with $a=180$, $b=2$, $p=q=0.7$. 95\% confidence intervals are in brackets. The sample size is $n=500$.}
%	\bigskip
	\begin{tabular}{c|ccc|ccc}
		\midrule
& \multicolumn{3}{c}{VaR} & KS & AD & BIC\\
\midrule
		& 0.95  & 0.99  & 0.995 & & & \\
		\midrule
		\multirow{2}{*}{Static mix.} & 1256.02 & 3962.18 & 6555.69 & \multirow{2}{*}{0.704} & \multirow{2}{*}{0.735} & \multirow{2}{*}{6801.73} \\
		& [1037.66,1603.11] & [2202.45,6308.31] & [2965.99,11\,318.27] &  & & \\
\midrule
		\multirow{2}{*}{Case 1} & 1824.26 & 5693.55 & 9151.98 & \multirow{2}{*}{$<0.001$} & \multirow{2}{*}{$<0.001$}  & \multirow{2}{*}{6872.14}\\
		& [1437.77,2681.18] & [2458.05,9771.98] & [5413.29,19\,007.93] & & & \\
\midrule
		\multirow{2}{*}{Case 2} & 1449.17 & 4525.31 & 7371.63 & \multirow{2}{*}{0.004} & \multirow{2}{*}{0.003}  & \multirow{2}{*}{6807.95} \\
		& [1088.67,2060.20] & [2710.36,8209.61] & [3752.41,16\,120.79] &  & & \\
\midrule
		True & 1238.67 & 3932.81 & 6456.66 & - & - & \\
		\midrule
	\end{tabular}
	\label{tab:GB2}
	%	SimGB2.R
\end{table}

Also in this case, the static mixture fit is excellent, whereas the KS and AD tests suggest that the remaining two distributions do not model the data appropriately. The BIC also favors the static mixture. Accordingly, the static mixture VaR is in line with the observed quantile at all levels, and displays a smaller variability.

\subsection{Computational issues}
\label{sec:comp}

The EM algorithm of Sect. \ref{sec:EM} is implemented in the \texttt{lognGPD} \texttt{R} package \citep{CRAN:lognGPD}. In addition to estimation, calculation of bootstrap standard errors is supported and carried out via parallel computing. Functions for random number simulation and density evaluation are also available.

To assess the computational burden of the three approaches, we sample and estimate each of the three models considered so far, i.e. the static lognormal-GPD mixture, the composite lognormal-Pareto (Case 1) and the dynamic lognormal-GPD (Case 2). We use all methods in all setups, hence each of them is correctly-specified in turn. The sample size is $n=500$. 

\begin{table}[htbp]
	\centering
	\caption{Average computing times (in seconds) based on 1000 replications. The parameters of the true static mixture DGP are $p=0.9$, $\mu = 0$, $\sigma = 0.5$, $\xi = 0.5$, $\beta = 3.5$), of the true composite lognormal-Pareto (Case 1) are $\mu=0, \sigma = 0.5$, $x_{min} = 5$, $\alpha = 2$, and of the the true lognormal-GPD dynamic mixture (Case 2) are $\mu_c=1$, $\tau = 2$, $\mu=0$, $\sigma = 0.5$, $\xi = 0.5$, $\beta = 3.5$.}
%	\bigskip
	\begin{tabular}{c|ccc}
		\midrule
True DGP				& Static mix. & Composite (Case 1) & Dynamic mix. (Case 2) \\
\midrule
Static mix.  			& 0.145  & 0.189  & 0.173 \\
Composite (Case 1) 		&  0.085  & 0.015  & 0.107 \\
Dynamic mix. (Case 2) 	& 58.953  & 64.106  & 40.966 \\
\midrule
	\end{tabular}
\label{tab:times}
	%	SimGB2.R
\end{table}
The outcomes in Table \ref{tab:times} suggest that the EM-based approach is not as fast as MLE estimation of the lognormal-Pareto distribution, which is not surprising, given that the EM algorithm is known to be rather slow (see, e.g., \citealp[Section 3.9]{mcl08}). However, it is much faster than MLE in Case 2.

\section{Empirical analysis}
\label{sec:emp}

In this section we carry out two real-data analyses: the two datasets are related to automobile claims and to operational losses, respectively. In both cases we fit the static lognormal-GPD mixture as well as the pure lognormal and pure GPD distributions. Non-parametric bootstrap is used for estimating standard errors and confidence intervals: we draw with replacement $B=1000$ bootstrap samples of size $n$ (the observed sample size) and evaluate empirical quantities of the resulting bootstrap distribution.

\subsection{Automobile insurance claims}
\label{sec:app1}

We analyze the distribution of claims experience from a large Midwestern (US) property and casualty insurer for private passenger automobile insurance. The data are the amounts, in US dollars, paid on closed claims, and are publicly available in the \texttt{AutoClaims} dataset of the \texttt{R} package \texttt{insuranceData}. The sample size is $n=6773$.

%\FloatBarrier

\begin{figure}
	\begin{center}
		\includegraphics[width=11cm]{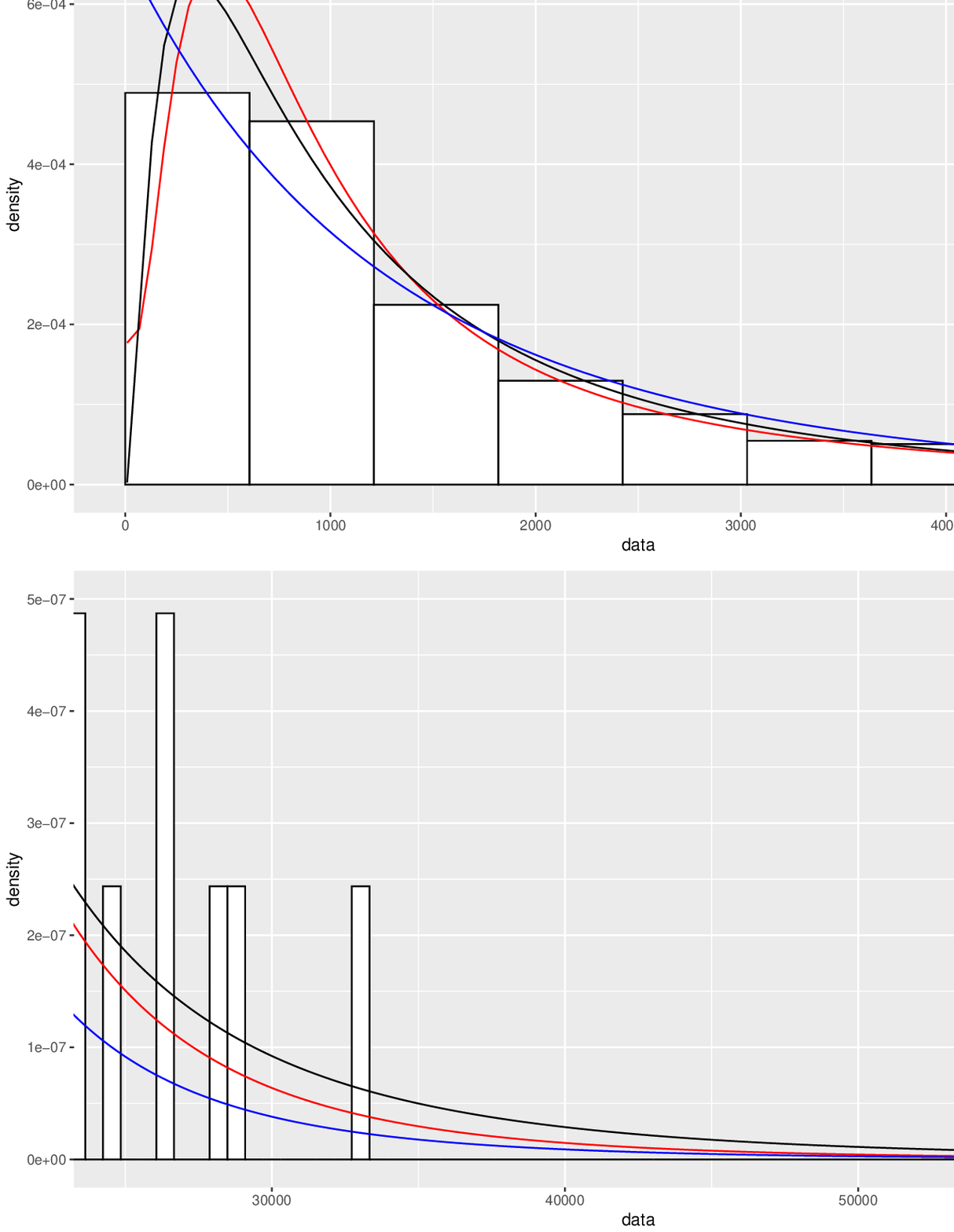}
		\caption{Histogram of the AutoClaims data, with superimposed the static mixture density (``Mixture''), the pure lognormal (``Lognormal'') and the pure GPD (``GPD''). Top panel: zoom on the distribution body ($x<5000$). Bottom panel: zoom on the distribution tail ($x>25\,000$). The bin width is the same in both panels.}
		\label{fig:threedens}
		% mainApp.R
	\end{center}
\end{figure}

% The histogram of the data, with superimposed the estimated density, is shown in Fig. \ref{fig:threedens}.
The upper panel of Figure \ref{fig:threedens} shows the histogram of the data with superimposed the estimated static mixture density: the fit looks quite good. Further support comes from Table \ref{tab:appGoF}, which reports the $p$-values of the KS and AD tests for the mixture, as well as well as for the pure lognormal and GPD: the GoF tests confirm that the estimated density is an appropriate model for the data\footnote{For both tests, we use the two-sample version, implemented in the \texttt{ks.test} and \texttt{ad.test} (from the \texttt{kSamples} package) \texttt{R} functions, respectively. The first sample is the observed sample, whereas the second one is simulated from the assumed distribution with estimated parameters. The $p$-value is computed from the asymptotic distribution of the test statistics, and double-checked via simulation (i.e., \texttt{simulate.p.value = TRUE} in \texttt{ks.test} and \texttt{method = ``simulated''} in \texttt{ad.test}). The outcomes are nearly identical.}. %Both the histogram and the goodness-of-Fit tests suggest a very good fit.

%\begin{table}[htbp]
%	\centering
%\caption{Automobile claims: observed $p$-value, average $p$-value and percentage of $p$-values smaller than 5\% (in the last two cases, across $B=500$ bootstrap replications), of the KS and AD tests for the estimated mixture, lognormal and GPD.}
%%	\bigskip
%	\begin{tabular}{c|ccc|ccc|cc}
%		& Mix. & Logn. & GPD & Mix. & Logn. & GPD& Mix. & Logn. \\
%		& \multicolumn{3}{c}{$p$-value} & \multicolumn{3}{c}{Average $p$-value} & \multicolumn{2}{c}{Perc.} \\
%		\hline
%		KS  & 0.787  & 0.044 & $<0.001$ & 0.787  & 0.044 & $<0.001$ & 0.787  & 0.044 \\
%		AD  & 0.829  & 0.021 & $<0.001$ & 0.787  & 0.044 & $<0.001$ & 0.787  & 0.044  \\
%		\hline
%	\end{tabular}
%	\label{tab:appGoF}
%\end{table}
\begin{table}[htbp]
	\centering
	\caption{Automobile claims: $p$-values of the GoF tests for the estimated static mixture, the pure lognormal and the pure GPD.}
	%	\bigskip
	\begin{tabular}{c|ccc}
		& Static mixture & Lognormal & GPD \\
		\midrule
		KS  & 0.787  & 0.044 & $<0.001$   \\
		AD  & 0.829  & 0.021 & $<0.001$ \\
		\midrule
	\end{tabular}
	\label{tab:appGoF}
\end{table}

\begin{figure}
	\begin{center}
		\includegraphics[width=12cm]{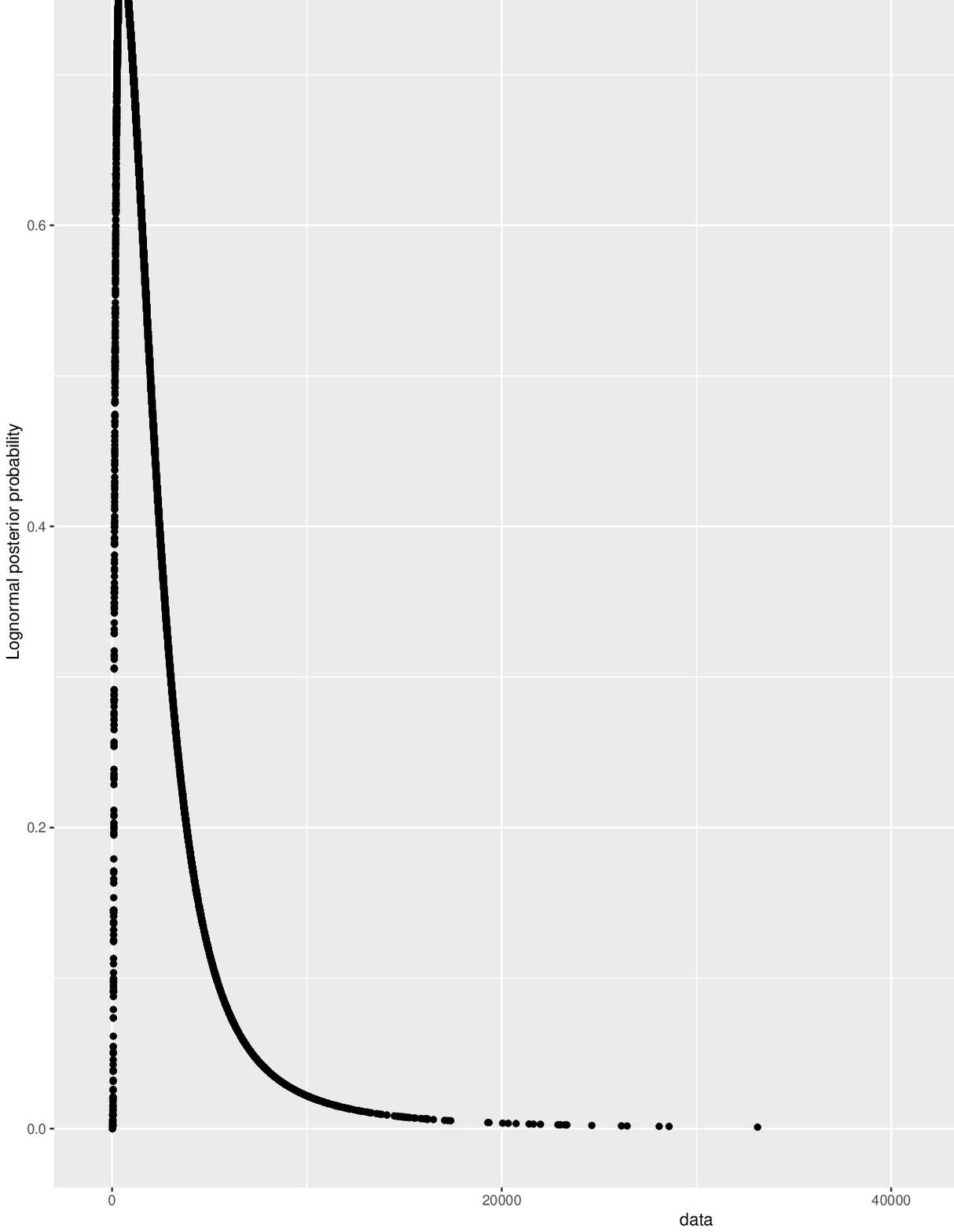}
		\caption{Scatterplot of the observations versus the posterior probabilities of belonging to the lognormal distribution, $\tau_{i1}$.}% Lower panel: zoom on the distribution body.}
	\label{fig:app2}
	% mainApp.R
\end{center}
\end{figure}

The middle and bottom panels of Figure \ref{fig:threedens} display the body and the tail of the estimated densities, as well as the lognormal and GPD densities estimated using all the observations. The latter is not a good fit to the smallest observations, a feature that is commonly observed for the GPD with $\xi>0$. For data in this range, the lognormal provides a better fit, and so does the static mixture, because most observations are lognormal, as made clear by the posterior probabilities in Figure \ref{fig:app2}. On the other hand, due to theoretical results in extreme value theory, the GPD is assumed to be best at modeling the tail, and the static mixture gets closer and closer to the GPD when we consider larger observations. 

%\FloatBarrier

Table \ref{tab:app1} reports parameter estimates, bootstrap standard errors based on 1000 replications, and 95\% confidence intervals. The outcomes imply an heavy-tailed distribution, since $\hat{\xi}>0$. The confidence intervals suggest a high precision: for example, the interval for $\xi$ goes from 0.102 to 0.205, thus confirming that the data are heavy-tailed.

\begin{table}[htbp]
	\centering
\caption{Parameter estimates, bootstrap standard errors and 95\% confidence intervals in the \texttt{AutoClaims} example.}
%	\bigskip
{\footnotesize	\begin{tabular}{ccccc}
		\midrule
		$p$  & $\mu$  & $\sigma$ & $\xi$  & $\beta$  \\
		\midrule
		0.567  &  6.676  &  0.752  &  0.156 & 2442.700 \\ 
		(0.038) &  (0.030) &  (0.034)  & (0.028) & (125.422)  \\
%{\tiny $[0.499, 0.645]$} & {\tiny $[6.618, 6.735]$} & {\tiny $[0.688, 0.820]$} & {\tiny $[0.102, 0.205]$} & {\tiny $[2240.414, 2725.608]$} & - & - \\
$[0.499, 0.645]$ & $[6.618, 6.735]$ & $[0.688, 0.820]$ & $[0.102, 0.205]$ & $[2240.414, 2725.608]$ \\
		\hline
	\end{tabular}}
	\label{tab:app1}
\end{table}

Figure \ref{fig:app2} is a scatterplot of the observations versus the posterior probabilities of belonging to the lognormal distribution, $\tau_{i1}$. 
For the bulk of the data, the posterior probability takes intermediate values, so that both distributions are likely to have generated them. In particular, for all the observations between $x_{(172)}$ and $x_{(5339)}$, $\tau_{i1}\in[0.40,0.780]$, where 0.780 is the largest value of $\tau_{i1}$ for this dataset. On the other hand, the largest observations belong to the GPD with probability very close to 1: the top 50 observations have $\tau_{i2}>0.99$, that is, are very likely to be generated by the GPD. Even though high posterior probabilities for extreme observations do not, by themselves, allow us to draw conclusions about clustering suitability, this suggests a possible use of the model for classification purposes.

Finally, we estimate VaR at three different levels. Since the GoF tests suggest that the lognormal and GPD distributions are not flexible enough, VaR is computed also via the composite lognormal-Pareto (Case 1) and the dynamic mixture (Case 2). %To assess variability, the VaR is recomputed with the estimated parameter values obtained in the 1000 non-parametric bootstrap replications employed for the bootstrap analysis in Table \ref{tab:app1}.

\begin{table}[htbp]
	\centering
	\caption{VaR estimates, bootstrap standard errors and 95\% confidence intervals in the \texttt{AutoClaims} example. For comparison purposes, the last line reports the quantiles of the observed data.}
	\begin{tabular}{cccc}
		\midrule
&		$\alpha=95\%$  & $\alpha=99\%$  & $\alpha=99.5\%$  \\
		\midrule
Static mix. & 6382.85 & 12\,540.60 & 15\,698.36 \\
&		(222.87) &  (682.73) & (1082.96)  \\
&		$[5950.74, 6834.79]$ & $[11\,250.84, 13\,952.80]$ & $[13\,633.93, 17\,959.04]$ \\
		\midrule
Logn.	& 6114.77 & 12\,668.93 & 16\,532.58 \\
&		(172.8277) &  (707.6939) & (1323.3880)  \\
&		$[5815.11, 6451.20]$ & $[11\,533.03, 14\,862.41]$ & $[14\,119.81, 19\,489.84]$ \\
\midrule
GPD	& 6084.06 & 11\,096.40 & 13\,790.20 \\
&		(139.81) & (445.68) & (727.97)  \\
&		$[5803.2, 6341.10]$ & $[10\,212.47, 11\,958.85]$ & $[12\,456.99, 15\,351.07]$ \\
\midrule
Case 1 & 6810.63   &   15\,048.99   &   20\,426.43 \\
& (545.06) &     (1906.43)     &  (3226.70)  \\
& $[5862.78, 7967.64]$ & $[11\,715.95, 19\,227.58]$ & $[15\,039.30, 27\,668.46]$ \\
		\midrule
Case 2 & 5470.00 & 11\,459.60 & 15\,036.81 \\
& (590.60) & (1587.31) & (2391.44)  \\
& $[4896.07, 6875.40]$ & $[9714.15, 15\,383.81]$ & $[12\,303.07, 21\,210.88]$ \\
\midrule
Empirical & 6737.81   &   12\,713.65   &   15\,024.10 \\
		\midrule
\end{tabular}
	\label{tab:VaRapp}
\end{table}
\FloatBarrier

From Table \ref{tab:VaRapp} we see that the static lognormal-GPD mixture and the dynamic Cauchy-lognormal-GPD mixture (Case 2) yield similar measures, but the former is more precise at high levels, as can be seen by comparing the estimated standard deviations and the width of the confidence intervals. %In addition, as seen in Section \ref{sec:comp}, the latter has a much heavier computational burden.
Notice also that both methods yield VaR measures in line with the corresponding quantiles of the observed data. On the other hand, the composite lognormal-Pareto distribution (Case 1) not only gives higher VaR estimates, but also a larger variability. The lognormal-based VaR is reasonably accurate, in line with the GoF tests for the lognormal distribution (see Table \ref{tab:appGoF}) being only marginally significant. Finally, the GPD underestimates the VaR: this may be due to the threshold being equal to zero, which results in a poor fit to the smallest observations.

\subsection{Operational risk}

\begin{figure}
	\begin{center}
		\includegraphics[width=11cm]{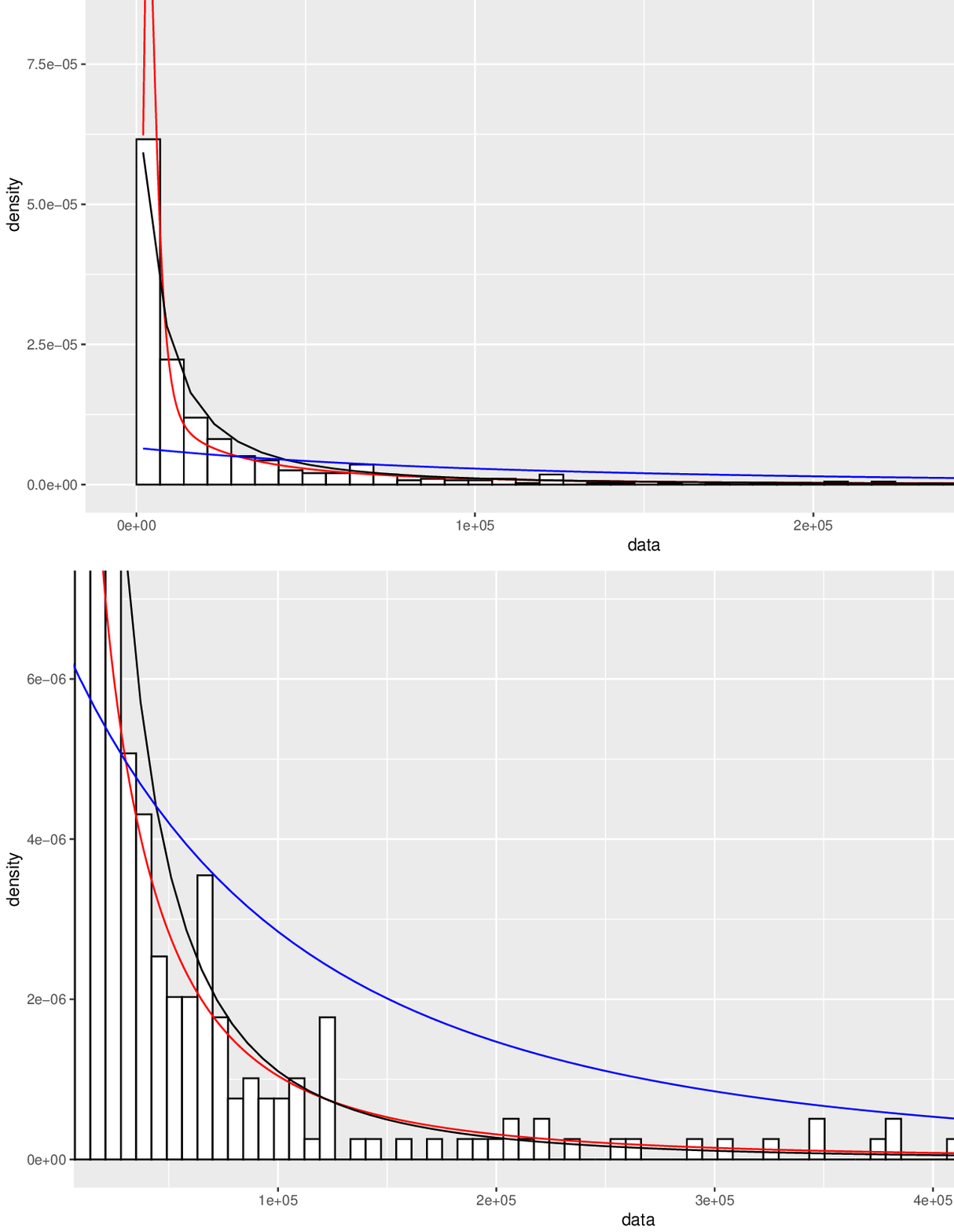}
		\caption{Histogram of the operational risk data, with superimposed the lognormal-GPD density, as well as the estimated lognormal and GPD densities. Top panel: zoom on the distribution body ($x<30\,000$). Bottom panel: zoom on the distribution tail ($30\,000<x<500\,000$). The 17 observations larger than 500\,000 Euro have been omitted for readability. The bin width is the same in both panels.}
		\label{fig:threedens2}
		% mainApp.R
	\end{center}
\end{figure}

According to \cite{basel04}, operational risk is the ``loss from failed internal processes, people, systems, or external events, covering areas like fraud, business disruption, and legal issues''.
In this section we model operational risk losses registered in the Unicredit Italian bank. In particular, we fit losses (in Euro) observed in the \emph{Internal Fraud} business line, from January 1, 2005 to June 30, 2014; the anonymized data are available as Supplementary File. A graphical representation is given in Fig. \ref{fig:threedens2}. Also in this case, the distribution is right-skewed and leptokurtic: the empirical skewness and kurtosis are equal to 12.57 and 182.82, respectively. For readability, the upper panel displays only the observations smaller than 30\,000, along with the densities estimated via MLE and lognormal NCE. Similarly, the lower panel refers to losses between 30\,000 and 500\,000. For readability, in the latter plot we had to omit 17 observations larger than 500\,000 (the largest one is equal to 7\,000\,000 Euro); for the same reason, the histogram of the whole dataset is omitted.
Also in this dataset, the GPD fit to the body is poor, probably because of a very large large value $\hat{\xi}$ (see Table \ref{tab:app2} below), related to the presence of a few very large losses. For data in this range, the lognormal provides a better fit. %On the other hand, due to theoretical results in extreme value theory, the GPD is assumed to be best at modeling the tail, and the mixture gets closer and closer to the GPD when we consider larger observations. 

To assess the mixture GoF, Table \ref{tab:app2GoF} displays the outcomes of the GoF tests for the mixture, the lognormal and the GPD: the GoF tests clearly tell that the estimated mixture density is an appropriate model for the data, unlike the lognormal and GPD.

\begin{table}[htbp]
	\centering
	\caption{Operational risk: $p$-values of the GoF tests for the estimated mixture, lognormal and GPD.}
	%	\bigskip
	\begin{tabular}{c|ccc}
		& Static mixture & Lognormal & GPD \\
		\hline
		KS  & 0.200  & $<0.001$ & $<0.001$   \\
		AD  & 0.081  & $<0.001$ & $<0.001$ \\
		\hline
	\end{tabular}
	\label{tab:app2GoF}
\end{table}

Table \ref{tab:app2} reports point estimates and 95\% confidence intervals. The large positive value of $\hat{\xi}$ suggests a very heavy tail. With respect to the previous application, both the standard errors and the confidence intervals reveal less precise estimates, a result that is not surprising, given the much smaller sample size: however, also in this case, the interval for $\xi$ allows us to confirm that the distribution is heavy-tailed.

\begin{table}[htbp]
	\centering
	\caption{Parameter estimates, bootstrap standard errors and 95\% confidence intervals in the operational risk example.}
	%	\bigskip
	{\footnotesize	\begin{tabular}{ccccc}
			\hline
			$p$  & $\mu$  & $\sigma$ & $\xi$  & $\beta$  \\
			\hline
			0.430   &  8.401  &   0.491   &  0.831 & 31076.944  \\ 
			(0.066) & (0.118) & (0.084) & (0.088) & (5502.945)  \\
			$[0.344, 0.599]$ & $[8.248, 8.724]$ & $[0.388, 0.710]$ & $[0.649, 0.997]$ & $[25\,747.03, 48\,091.6]$ \\
			\hline
	\end{tabular}}
	\label{tab:app2}
\end{table}

To conclude the analysis, we estimate VaR with the same approach used in Section \ref{sec:app1}. Table \ref{tab:VaRapp2} displays the results.

\begin{table}[htbp]
	\centering
	\caption{VaR estimates, bootstrap standard errors and 95\% confidence intervals in the operational risk example. For comparison purposes, the last line reports the quantiles of the observed data.}
	%	\bigskip
	\begin{tabular}{cccc}
		\midrule
		&		$\alpha=95\%$  & $\alpha=99\%$  & $\alpha=99.5\%$  \\
		\midrule
		Static mix. & 249\,543.6 & 1\,067\,999.4 & 2\,004\,661.8 \\
		&		(61\,688.09) & (529\,322.29) & (1\,401\,654.06)  \\
		&		$[155\,586.7, 396\,723.0]$ & $[458\,207.3, 2\,344\,718.2]$ & $[604\,048.3, 5\,410\,040.0]$ \\
		\midrule
		Logn.	& 162\,069.13 & 450\,421.15 & 642\,360.86 \\
&		(20\,547.63) & (102\,242.96) & (200\,289.43)  \\
&		$[127\,608.4, 204\,094.0]$ & $[297\,247.3, 684\,216.8]$ & $[371\,838.9, 1\,143\,820.5]$ \\
\midrule
		GPD	& 537\,766.36 & 1\,290\,757.84 & 1\,830\,817.89 \\
&		(62\,332.55) & (272\,792.24) & (533\,200.10)  \\
&		$[425\,682.3, 672\,890.2]$ & $[881\,114.9, 1\,921\,697.8]$ & $[1\,102\,718.1, 3\,100\,422.6]$ \\
\midrule
		Case 1 & 250\,314.60 &  1\,564\,027.72 &  2\,910\,657.39 \\
		& (80.391.24) &     (806\,004.43)     &  (2\,326\,771.70)  \\
		& $[38\,652.78, 739\,567.64]$ & $[311\,749.92, 3\,016.59]$ & $[883\,576.24, 7\,501\,942.79]$ \\
		\midrule
		Case 2 & 198\,712.2 & 915\,193.5 & 1\,749\,429.6 \\
		& (59\,699.44) & (526\,022.86) & (1\,257\,898.64)  \\
		& $[115\,109.9, 346\,263.0]$ & $[370\,096.6, 2\,148\,516.3]$ & $[544\,911, 5\,452\,726]$ \\
		\midrule
		Empirical & 287\,024 & 1\,256\,662 & 1\,719\,884  \\
		\midrule
	\end{tabular}
	\label{tab:VaRapp2}
\end{table}

Similarly to the previous application, the static mixture yields estimated VaRs closer to the empirical quantile and characterized by a smaller standard error with respect to the dynamic mixture (Case 2). On the other hand, the VaRs based on the composite lognormal-Pareto distribution (Case 1) seem to be overestimated. Finally, the lognormal VaR is clearly too low, and the GPD VaR is rather good at levels 99 and 99.5\%, but quite large when $\alpha=95\%$.

\section{Conclusion}
\label{sec:concl}

In this paper we have proposed a flexible unsupervised lognormal-GPD mixture for skewed and heavy-tailed distributions. A maximum likelihood estimation procedure based on the EM algorithm is developed and its properties are studied via simulation experiments.

With respect to competing models, the advantages range from a simple mixture structure to a reliable estimation procedure characterized by a low computational burden. %Moreover, the method is more interpretable than the two alternative we have considered.
By ``simple mixture structure'' we mean that it is a two-population mixture with a continuous density, without the need of constraints, unlike the lognormal-Pareto composite model developed by \cite{scoll07}; moreover, with respect to the dynamic mixture proposed by \cite{fri02}, the density is normalized, so that it is not necessary to tackle the additional problem of evaluating the normalization constant.

In terms of flexibility, the static lognormal-GPD mixture yields excellent results when used for fitting data in mis-specified setups, and allows the investigator to estimate VaR measures quite precisely. In particular, the VaR computed via the static lognormal-GPD mixture model is less variable than via the lognormal-GPD dynamic mixture.

In principle, the proposed approach may be extended to the case of left-censored (or truncated) data. In such a setup, the M-steps for the lognormal parameters would not be in closed form either, but it should be possible to use a nested EM-algorithm \citep{vand00}: at the $t$-th iteration of the EM algorithm presented in Section \ref{sec:EM}, the lognormal parameters are estimated by means of an EM algorithm for MLE of the censored or truncated lognormal distribution \citep{bee06}.

Even though the main purpose of this paper is the development and estimation of a model for skewed and heavy-tailed data, there are setups where classification is the main purpose, i.e. where the main goal is identifying which observations are lognormal and which are GPD. Since one of the byproducts of the EM algorithm are posterior probability of the observations, our approach is very well suited for this aim.

From the point of view of real-data analysis, in Section \ref{sec:emp} we have limited ourselves to comparisons with simple approaches, based on the pure lognormal or GPD distribution. Even though other composite and contaminated distributions may not be well-suited to modeling insurance and operational loss data such as those used in the present paper, it may be of interest to empirically assess their goodness-of-fit, especially if working in applications outside the insurance analytics and economic loss settings.

\section*{Declarations}

\begin{itemize}
	\item Conflict of interest/Competing interests: the authors declare no conflict of interest and no competing interests.
	\item Code availability: All the methods developed in this paper are implemented in the \texttt{lognGPD} \texttt{R} package \citep{CRAN:lognGPD}. The anonymized data employed in the operational risk applications, as well as the codes for the two real-data analyses, are available as supplementary files.
\end{itemize}

\bibliography{c:/users/marco.bee/dropbox/biblio/probability,c:/users/marco.bee/dropbox/biblio/risk,c:/users/marco.bee/dropbox/biblio/actuarial,c:/users/marco.bee/dropbox/biblio/algorithms,c:/users/marco.bee/dropbox/biblio/distr,c:/users/marco.bee/dropbox/biblio/math_stat,c:/users/marco.bee/dropbox/biblio/likelihood,c:/users/marco.bee/dropbox/biblio/mixtures,c:/users/marco.bee/dropbox/biblio/biblioEE,c:/users/marco.bee/dropbox/biblio/biblioDacci}

\end{document}